\documentclass[12pt]{iopart}
\usepackage{iopams}  
\usepackage[pdftex]{graphicx}
\usepackage[colorlinks=true,citecolor=blue,linkcolor=blue]{hyperref}
\usepackage[utf8]{inputenc}
\usepackage{units}
\usepackage[usenames]{color}
\usepackage{pdfpages}

\renewcommand{\vec}[1]{\ensuremath{\boldsymbol{#1}}}

\newcommand{\rdep}{\left(\vec{r}\right)}
\newcommand{\erf}{\mathrm{erf}\!}

\begin{document}

\title[Band gaps of 2D materials from the LMBJ potential]{Accurate electronic band gaps of two-dimensional materials from the local modified Becke-Johnson potential}

\author{Tom\'{a}\v{s} Rauch$^1$, Miguel A. L. Marques$^{2,3}$, Silvana Botti$^{1,3}$}

\address{$^1$ Institut für Festkörpertheorie und -optik, Friedrich-Schiller-Universität Jena, Max-Wien-Platz 1, 07743 Jena, Germany}
\address{$^2$ Institut für Physik, Martin-Luther-Universität Halle-Wittenberg, 06120 Halle/Saale, Germany}
\address{$^3$ European Theoretical Spectroscopy Facility}


\ead{\mailto{tomas.rauch@uni-jena.de}}

\begin{abstract}
The electronic band structures of two-dimensional materials are significantly different from those of their bulk counterparts, due to quantum confinement and strong modifications of electronic screening. An accurate determination of electronic states is a prerequisite to design electronic or optoelectronic applications of two-dimensional materials, however, most of the theoretical methods we have available to compute band gaps are either inaccurate, computationally expensive, or only applicable to bulk systems. Here we show that reliable band structures of nanostructured systems can now be efficiently calculated using density-functional theory with the local modified Becke-Johnson exchange-correlation functional that we recently proposed. After re-optimizing the parameters of this functional specifically for two-dimensional materials, we show, for a test set of almost 300 systems, that the obtained band gaps are of comparable quality as those obtained using the best hybrid functionals, but at a very reduced computational cost. These results open the way for accurate high-throughput studies of band-structures of two-dimensional materials and for the study of van der Waals heterostructures with large unit cells.
\end{abstract}

%
%
%
%
%

\section{Introduction}
Since the discovery of graphene~\cite{Novoselov2004}, we have witnessed a huge interest in two-dimensional (2D) materials. This is motivated not only by the many potential applications of these nanostructured systems~\cite{Kerativitayana2015,Lee2018,Hu2018,Briggs2019,Zavabeti2020}, but also by the many fundamental physics questions pertaining electronic excitations in low-dimensionality~\cite{Bhimanapati2015,Ferrari2015}. Many different 2D materials have been synthesized, and many more have been predicted and theoretically investigated for the past years~\cite{Choudhary2017,Haastrup_2018,Zhou2019,Mounet2018}. In fact, the current availability of supercomputers and modern tools based on density-functional theory~\cite{Hohenberg1964,Kohn1965} (DFT) has allowed thorough investigations of a myriad of hypothetical 2D crystals. This new strategy, of computationally accelerated discovery of new materials, has emerged as the most powerful and cost effective research paradigm within materials science~\cite{Curtarolo2013high}.
Predicted 2D materials can then be characterized, again using computational methods, selected based on their properties, and then proposed for specific technological applications. The results of these high-throughput calculations are usually accessible from online databases. The most prominent for 2D materials are the JARVIS-DFT database~\cite{Choudhary2017,JarvisDFT} (1356 materials), the Computational 2D Materials Database~\cite{Haastrup_2018,C2DB} ($\backsim$ 4000 materials), 2DMatPedia~\cite{Zhou2019,2DMatPedia} ($\backsim$ 6000 materials), and the Materials Cloud Archive~\cite{Mounet2018,MaterialsCloud} ($\backsim$ 1000 materials).

Some of the most exciting prospective applications for 2D materials reside currently in the fields of electronics and optoelectronics~\cite{Kang2020}. In fact, several 2D semiconducting materials with large band gaps and high carrier mobilities have recently attracted tremendous attention. This includes not only graphene and transition metal dichalcogenides, but also monolayers like black phosphorus~\cite{Castellanos2015}, that exhibits hole mobilities as large as $\backsim 1000$~cm$^2$V$^{-1}$s$^{-1}$. Furthermore, a big advantage of novel 2D semiconductors is that their electronic and transport properties~\cite{Sangwan2018} can be engineered by modifying their thickness or applying external fields, leading to unprecedented physical properties that are not observed in bulk semiconductors (for example, the tunability of carriers~\cite{Cui2015multi} or controllable valley physics~\cite{Ye2017optical}). Devices can be easily built by stacking mono- or few-layer semiconductors and metals to obtain van der Waals heterostructures~\cite{geim2013van}.

The key physical property to evaluate electronic and optoelectronic applications is the band structure, and in particular the band gap. Of course, for the high-throughput calculations mentioned above to be meaningful, our theoretical methods have to be able to provide an accurate estimation of such properties. These methods should moreover be computationally efficient to allow for a time-effective screening of large classes of materials.

DFT is often the method of choice to calculate the fundamental band gaps in both three-dimensional (3D) and 2D worlds. Unfortunately, the quality of the results obtained with DFT depends strongly on the choice of the approximation to the exchange-correlation (XC) functional, unavoidable in any practical implementation of the theory. Hundreds of approximate XC functionals have been proposed in the literature~\cite{Lehtola2018,Marques2012} and the reliability of some of them for band gap calculations was evaluated in several studies for 3D materials~\cite{tranJ.Phys.Chem.A2017,coskunJ.Chem.TheoryComput.2016,garzaJ.Phys.Chem.Lett.2016,peveratiJ.Chem.Phys.2012,crowleyJ.Phys.Chem.Lett.2016,tranPhys.Rev.Mater.2018,Borlido2019,Borlido2020}. The modified Becke-Johnson (MBJ)~\cite{Tran2009}, the 2006 hybrid from Heyd–Scuseria–Ernzerhof (HSE06)~\cite{HSE03,HSE06}, and the high-local exchange (HLE16)~\cite{Verma2017} XC functionals were identified as those with the smallest error for band gaps. In particular, the MBJ combined the best performance overall with a computational effort orders of magnitude smaller than hybrid functionals~\cite{Borlido2020}. Note that, even though all of the three functionals clearly outperform the Perdew-Burke-Ernzerhof (PBE) functional~\cite{Perdew1996}, the latter is still largely used in standard calculations of band structures.

In spite of all its qualities, the MBJ potential is unsuitable for being used for 2D systems. The problem resides in an integral that enters in the definition of the exchange potential and that is only defined for 3D bulk systems.
Recently, we proposed an improvement of the MBJ potential, the local MBJ (LMBJ) potential~\cite{Rauch2020}, that resolves this weakness. In this work, we aim at the following two goals: (i)~reoptimizing the two new parameters of the LMBJ (i.e., beside the parameters already present in the MBJ potential) to minimize the error for band gaps of 2D materials; and (ii)~evaluating the performance of the LMBJ potential for a large data set of 2D systems. In this way, we verify if the local version of the of the MBJ potential preserves the good description of band gaps also for lower-dimensional and heterogeneous systems. 

In benchmark studies it is common to evaluate errors of theoretical calculations with respect to known experimental data. Unfortunately, accurate experimental data on band gaps of 2D materials are still scarce. Therefore, we chose to measure the error of our band gap calculations with respect to the best accessible theoretical data. These are results of many-body $G_0W_0$ calculations, which are stored for hundreds of 2D materials in the C2DB database~\cite{Haastrup_2018,C2DB}. We filtered out 298 stable, nonmagnetic materials with known $G_0W_0$ band gaps. This is enough data to fit the parameters of the LMBJ potential and to evaluate its performance. We should note, however, that $G_0W_0$ band gaps have a non-negligible error~\cite{Setten2017} when compared to experimental data, and that it is well known that this approximation can catastrophically fail for $d$-electron systems~\cite{Aryasetiawan1995,Continenza1999,Bruneval2006}. This means that it is not uncommon for 3D materials that the MBJ gap from Kohn-Sham DFT calculations is closer to experiment than the $G_0W_0$  band gap. This fact will of course be kept in mind when discussing our results.

\section{Methods}
\subsection{Local modified Becke-Johnson potential}
The original Becke-Johnson (BJ) potential~\cite{Becke2006} is a semilocal meta-generalized gradient approximation potential for the exchange part of the exchange-correlation potential, and it was modified by Tran and Blaha to obtain electronic band gaps of three-dimensional semiconductors very close to experimental values~\cite{Tran2009}. It has the form
\begin{equation}
    \label{eq:MBJ_def}
    v^{\mathrm{MBJ}}_{x}\rdep = c v^{\mathrm{BR}}_{x}\rdep + \left(3 c - 2\right) \frac{1}{\pi} \sqrt{\frac{5}{12}}\sqrt{\frac{2 t \rdep}{\rho \rdep}} \,,
\end{equation}
with the electronic density $\rho\rdep = \sum^{N}_{i} \left|\psi_{i}\rdep\right|^2$, the kinetic-energy density $t\rdep = \sum^{N}_{i} \nabla\psi^{*}_{i} \cdot \nabla\psi_{i}$ and the Becke-Russel (BR) exchange potential~\cite{Becke1989}
\begin{equation}
    \label{eq:BR_def}
    v^{\mathrm{BR}}_{x}\rdep = -\frac{1}{b\rdep}\left[ 1-e^{-x\rdep} - \frac{1}{2}x\rdep e^{-x\rdep} \right].
\end{equation}
$x\rdep$ is calculated from $\rho\rdep$ and its spatial gradient and Laplacian, and \mbox{$b\rdep = \sqrt[3]{x^3 e^{-x}/(8\pi\rho\rdep)}$}.
The crucial part of the MBJ potential is the $c$ parameter, which is calculated self-consistently as
\begin{equation}
    \label{eq:c_def}
    c = \alpha + \beta \bar{g}^\epsilon
\end{equation}
with
\begin{equation}
    \label{eq:g_def}
    \bar{g} = \frac{1}{V_{\mathrm{cell}}} \int_{\mathrm{cell}} d^{3}r\ \frac{\left|\nabla \rho\rdep\right|}{\rho\rdep}.
\end{equation}
$\alpha$ and $\beta$ are fitted parameters obtained by minimizing the average error of band gaps calculated for a set of semiconductors using the MBJ potential with respect to experimental values. In our work we will use the parameters $\alpha=0.488$ and $\beta=\unit[0.5]{bohr}$ together with $\epsilon = 1$ obtained during a later re-optimization of the potential~\cite{Koller2012}. Tran and Blaha defined the full XC potential by adding to Eq.~\ref{eq:MBJ_def} a correlation term in the local-density approximation: this has become the standard procedure to build the XC MBJ potential.

The fact that $g\rdep=\left|\nabla\rho\rdep\right|/\rho\rdep$ is averaged in the unit cell hinders the application of the MBJ potential to systems with reduced dimensionality (such as surfaces, mono- and few-layers, or molecules), as the value of $c$ becomes dependent on the fraction of vacuum in the supercell. The functional is for the same reason inadequate to study heterostructures or  other strongly inhomogeneous systems. To address this problem, we have recently introduced the LMBJ potential~\cite{Rauch2020}, a generalization of the MBJ exchange potential~\cite{Tran2009} that is identical to the parent functional for bulk crystals, but it can also be meaningfully used for inhomogeneous and nanostuctured materials. In fact, the mixing parameter of LMBJ potential
\begin{equation}
    \label{eq:c_loc}
    c\rdep = \alpha + \beta \bar{g}\rdep
\end{equation}
is a local function of $\vec{r}$  since $\bar{g}$ becomes a coordinate-dependent, locally averaged function
\begin{equation}
    \label{eq:g_loc}
    \bar{g}\rdep = \frac{1}{\left(2\pi\sigma^2\right)^{3/2}} \int d^3 r'\ g\left(\vec{r}'\right)\  e^{-\frac{\left|\vec{r}-\vec{r}'\right|^2}{2\sigma^2}} \,.
\end{equation}
The smearing parameter $\sigma$ in Eq.~\ref{eq:g_loc}  has a clear physical meaning, as it controls the localized volume over which $g\rdep$ is averaged. To guarantee a proper treatment of vacuum regions in the unit cell, i.e. to impose the correct asymptotic limit to the XC potential for large values of $r$, we further modified $g\rdep$, introducing the threshold density $\rho_{\mathrm{th}}$:
\begin{equation}
    \label{eq:g_vac}
    g\rdep = \frac{1-\alpha}{\beta}\left[1 - \erf\left(\frac{\rho\rdep}{\rho_{\mathrm{th}}}\right) \right]
    + \frac{\left|\nabla \rho\rdep\right|}{\rho\rdep} \erf\left(\frac{\rho\rdep}{\rho_{\mathrm{th}}}\right).
\end{equation}
 The threshold density $\rho_{\mathrm{th}}$ defines the value of the electronic density below which the LMBJ potential must be equal to the correct asymptotic limit in vacuum. In Ref.~\cite{Rauch2020} the two additional parameters of the LMBJ potential were fixed using physical arguments and it was shown that, for $\sigma$ = \unit[3.78]{bohr} = \unit[2]{\AA} and the threshold Wigner-Seitz radius $r^{\mathrm{th}}_s = (3/4\pi\rho_{\mathrm{th}})^{(1/3)} =$~\unit[5]{bohr}, the LMBJ potential yields band gap values of 3D materials almost identical to those of the MBJ potential. It is reasonable to expect that this remains true for $\sigma > $ \unit[3.78]{bohr}, as increasing the integration volume in Eq.~\ref{eq:g_loc} make us approach the definition of the MBJ potential. We also tested that $r^{\mathrm{th}}_s \in \left(3.0,7.0\right)$~bohr is a safe range for the threshold Wigner-Seitz radius. Given these ranges, the first aim of this work is to fit the two parameters to optimal values, so that we minimize the error in the calculation of band gaps of 2D semiconductors. Keeping the good quality of the original MBJ potential for 3D semiconductors is the obvious constraint to impose during this minimization procedure.

\subsection{The computational 2D materials database}
The computational 2D materials database (C2DB)~\cite{Haastrup_2018} collects various physical properties of more than 4000 2D materials calculated by DFT and many-body perturbation theory in the $G_0W_0$ approximation~\cite{Reining2018}. Since accurate experimental band gaps of 2D materials are so far unavailable, the currently best way to optimize and evaluate the predictive power of the LMBJ potential is to compare the LMBJ band gaps with those obtained by the most reliable theoretical method. Despite the issues with dielectric screening in 2D materials~\cite{Huser2013,Qiu2016,Rasmussen2016}, the state-of-the-art method for this class of systems is currently the $G_0W_0$ approximation~\cite{Rasmussen2015,Schmidt2017}. 

\subsection{Data set}
Starting from the assumption that $G_0W_0$ band gaps are currently the most reliable ones, we extracted from the C2DB the set of 2D materials that are stable (according to the stability criteria stated in the C2DB), non-magnetic, and for which the band gap has been calculated using $G_0W_0$. In this way we obtained 298 2D materials distributed among 12 different space groups. Out of these materials we chose 22 to be used for the optimization of the parameters $\sigma$ and $r^{\mathrm{th}}_s$, according to the following criteria: (i)~two materials were chosen from each space group (if possible); (ii)~as many chemical elements as possible should be represented in the chosen subset; (iii)~materials with a large range of band gaps should be included; and finally (iv)~experimentally known materials were given priority. The 22 2D materials selected according to these criteria are listed in Tab.~\ref{tab:chosen}.
%
\begin{table}[h!]
    \centering
    \begin{tabular}{c|c|ccc}
        spg & material & PBE band gap & HSE06 band gap & $G_0W_0$ band gap   \\
        \hline
        5 & S$_2$Tl$_2$ & 0.62 & 1.11 & 1.59\\
        5 & S$_2$Sn$_2$ & 1.94 & 2.56 & 3.25\\
        6 & Pb$_2$S$_2$ & 1.20 & 1.81 & 2.23\\
        6 & Cu$_2$S$_2$ & 0.62 & 1.31 & 1.58\\
        8 & Te$_2$Ti$_2$ & 0.11 & 0.15 & 0.27\\
        8 & Se$_2$Zn$_2$ & 1.61 & 2.64 & 3.46\\
        10 & Au$_2$O$_2$ & 0.18 & 0.87 & 0.97\\
        12 & Te$_2$Zr$_2$ & 0.21 & 0.25 & 0.45\\
        31 & Ge$_2$Se$_2$ & 1.12 & 1.56 & 1.86\\
        31 & S$_2$Si$_2$ & 1.42 & 2.14 & 2.88\\
        53 & As$_4$ & 0.83 & 1.32 & 1.80\\
        53 & P$_4$ & 0.90 & 1.51 & 2.03\\
        115 & CaF$_2$ & 6.45 & 8.63 & 11.37\\
        115 & CdBr$_2$ & 2.93 & 4.09 & 6.20\\
        129 & Cu$_2$Br$_2$ & 1.50 & 3.48 & 3.32\\
        129 & Cu$_2$Cl$_2$ & 1.45 & 3.66 & 3.45\\
        156 & ISSb & 1.22 & 1.78 & 2.26\\
        156 & CrSTe & 0.26 & 0.82 & 0.70\\
        164 & PtS$_2$ & 1.69 & 2.49 & 2.95\\
        164 & HfS$_2$ & 1.22 & 2.15 & 2.94\\
        187 & BN & 4.67 & 5.68 & 7.12\\
        187 & MoTe$_2$ & 0.93 & 1.37 & 1.56\\
    \end{tabular}
    \caption{Set of 22 2D materials chosen for the parameter optimization of the LMBJ potential. Band gaps are given in eV as stored in the C2DB~\cite{Haastrup_2018}. The column labelled ``spg'' is the space group of the crystal structure.}
    \label{tab:chosen}
\end{table}

The remaining 276 materials were used for the evaluation of the quality of the band gaps calculated by the LMBJ potential with the optimized parameters.
\begin{figure} 
  \centering
  \includegraphics[width = 0.99\columnwidth]{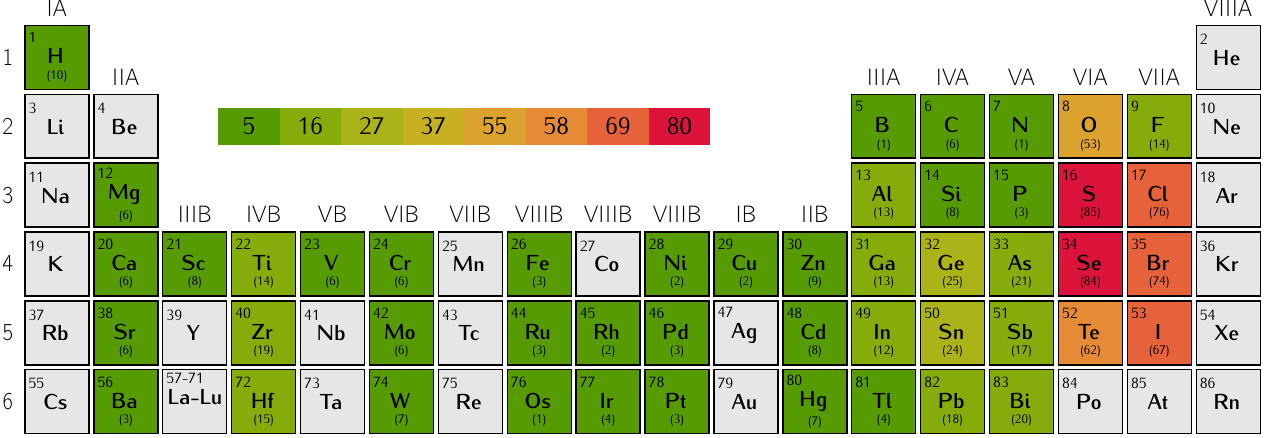}
  \caption{Frequency of elements in the evaluation data set. Elements indicated by gray boxes are not present in the data set.}
  \label{fig:fig1}
\end{figure}
The distribution of the individual elements among these 276 materials is shown in Fig.~\ref{fig:fig1}. Most of the periodic table is represented in the data set, but some parts are still missing, most prominently group IA elements (except hydrogen), noble gases, and lanthanides and actinides. On the contrary, the most represented elements are the nonmetals, in particular sulphur, selenium, and the halides.

The distribution of the $G_0W_0$ band gaps in the data set is the following: we find band gaps up to \unit[10.85]{eV}, with the majority of them lying in the interval between \unit[1]{eV} and \unit[4]{eV}. Out of the 276 materials, 161 are of $sp$-type containing only elements from groups IA, IIA, and IIA-VIIIA, and 115 are of $d$-type containing at least one transition metal. We will evaluate our results according to this subdivision of the data set and we denote the different sets ``all materials'', ``$sp$ materials'', and ``$d$ materials''.

\subsection{Computational details}
We performed all calculations using a custom version of the Vienna \textit{ab initio} simulation package (VASP) code~\cite{Kresse1996} with the projector-augmented-waves (PAW) method~\cite{Kresse1999}, where we implemented the LMBJ potential as described in the Supplementary Material of Ref.~\cite{Rauch2020}. Spin-orbit coupling was included self-consistently in all calculations.

All LMBJ calculations started from a converged calculation using the PBE functional~\cite{Perdew1996} and we always enforced non-magnetic solutions. The plane-wave cut-offs were taken from the values specified in the pseudopotentials distributed with VASP. All geometries were taken from the C2DB for consistency and two periodic 2D replicas were separated by \unit[15]{\AA} of vacuum. The densities of the \mbox{$\mathbf{k}$-point} meshes were dependent on the size of the Brillouin-zone and we set them to \unit[8.0 /]{\AA$^{-1}$} for all materials. Even though LMBJ calculations required a large number of iterations to reach convergence, as is known also from the MBJ implementation in VASP, all calculations converged for the 2D materials in the data sets.

\section{Results and discussion}
\subsection{Optimization of the parameters}
In the first step we optimized the parameters $\sigma$ and $r^{\mathrm{th}}_s$ of the LMBJ potential. For the selected set of 22 semiconductors we calculated the electronic band gaps using the LMBJ potential with $\sigma= \left\lbrace 2.0, 3.0, 4.0, 5.0, 6.0, 7.0  \right\rbrace$ \unit{\AA} and $r^{\mathrm{th}}_s = \left\lbrace 3.0, 5.0, 7.0  \right\rbrace$ \unit{bohr}. These values were chosen as they allow to recover the expected MBJ results for 3D materials.

For the optimization data set, our PBE results agree with those of the C2DB and the MAPE is in both cases larger than 50\%. For all the tested values for the additional parameters of the LMBJ potential we observe a large improvement over the PBE calculations, as the MAPE stays between 25\% and 35\%. The setting that gives the lowest MAPE ($\sigma= $\unit[4]{\AA} and $r^{\mathrm{th}}_s =$ \unit[7]{bohr}) predicts one false metal. Therefore, we choose for the LMBJ parameters the values $\sigma =$ \unit[4]{\AA} and $r^{\mathrm{th}}_s =$ \unit[5]{bohr} with the second best MAPE = 25\% and no false metals predicted.

\subsection{Evaluation using the control set}
\begin{figure} 
  \centering
  \includegraphics[width = 10cm]{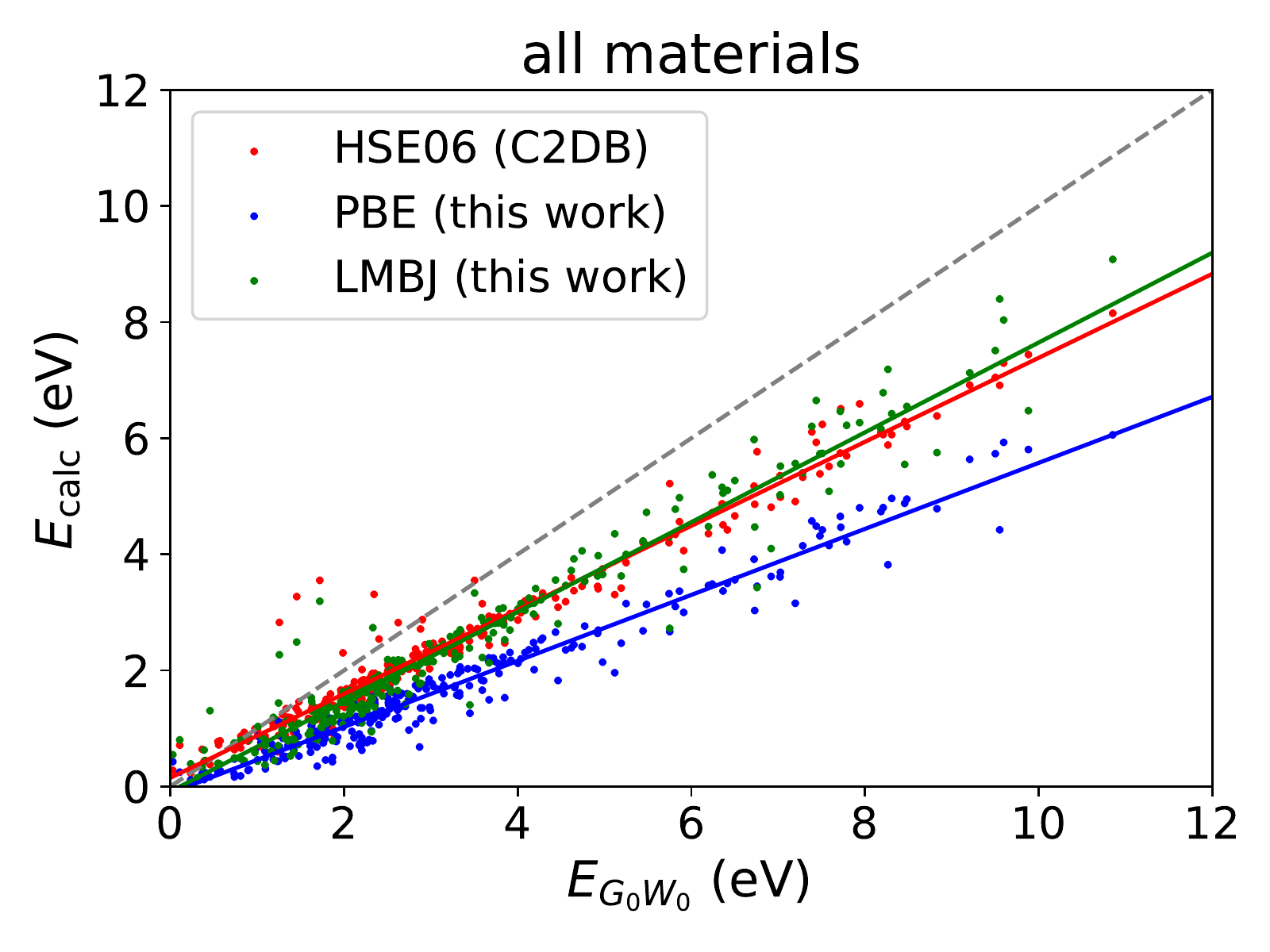}
  \caption{Calculated band gaps as a function of $G_0W_0$ (C2DB) band gaps. Full lines are linear fits ($y = ax+b$) to the respective data with $a$ and $b$ given in Tab.~\ref{tab:results}.}
  \label{fig:fig4}
\end{figure}

\begin{figure} 
  \centering
  \includegraphics[trim = {50, 0, 50, 0}, clip,width = 0.32\columnwidth]{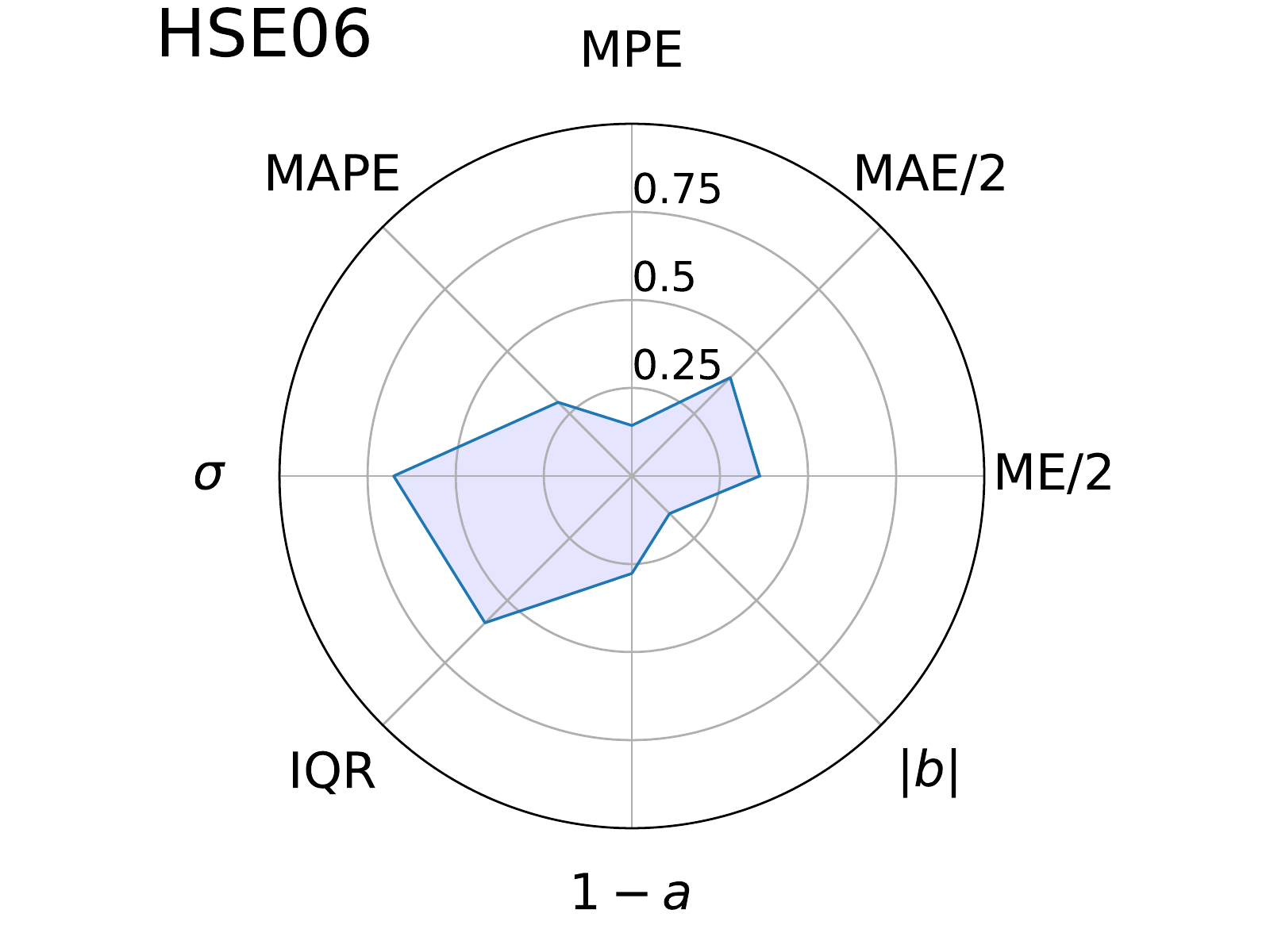}
  \includegraphics[trim = {50, 0, 50, 0}, clip,width = 0.32\columnwidth]{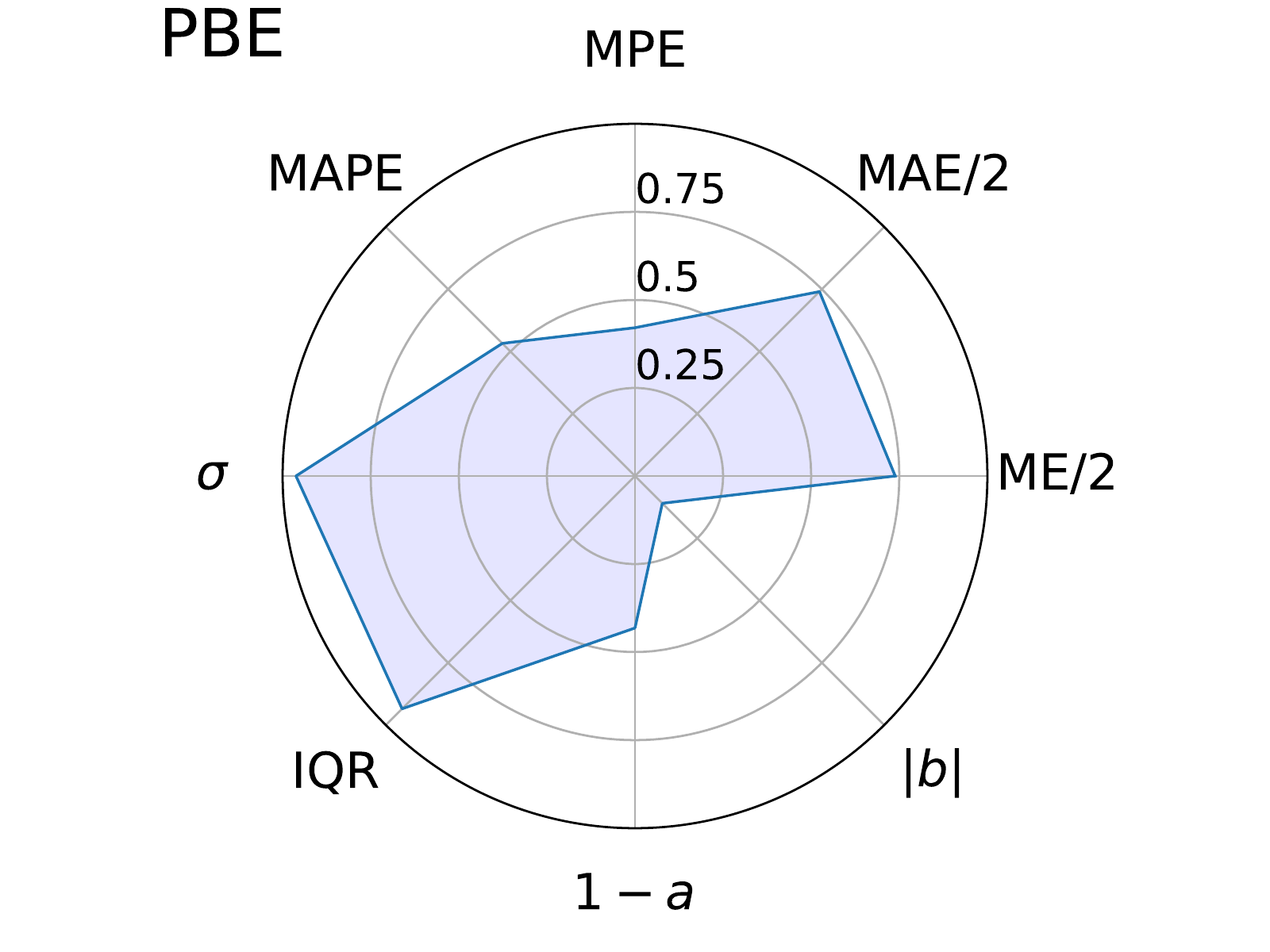}
  \includegraphics[trim = {50, 0, 50, 0}, clip,width = 0.32\columnwidth]{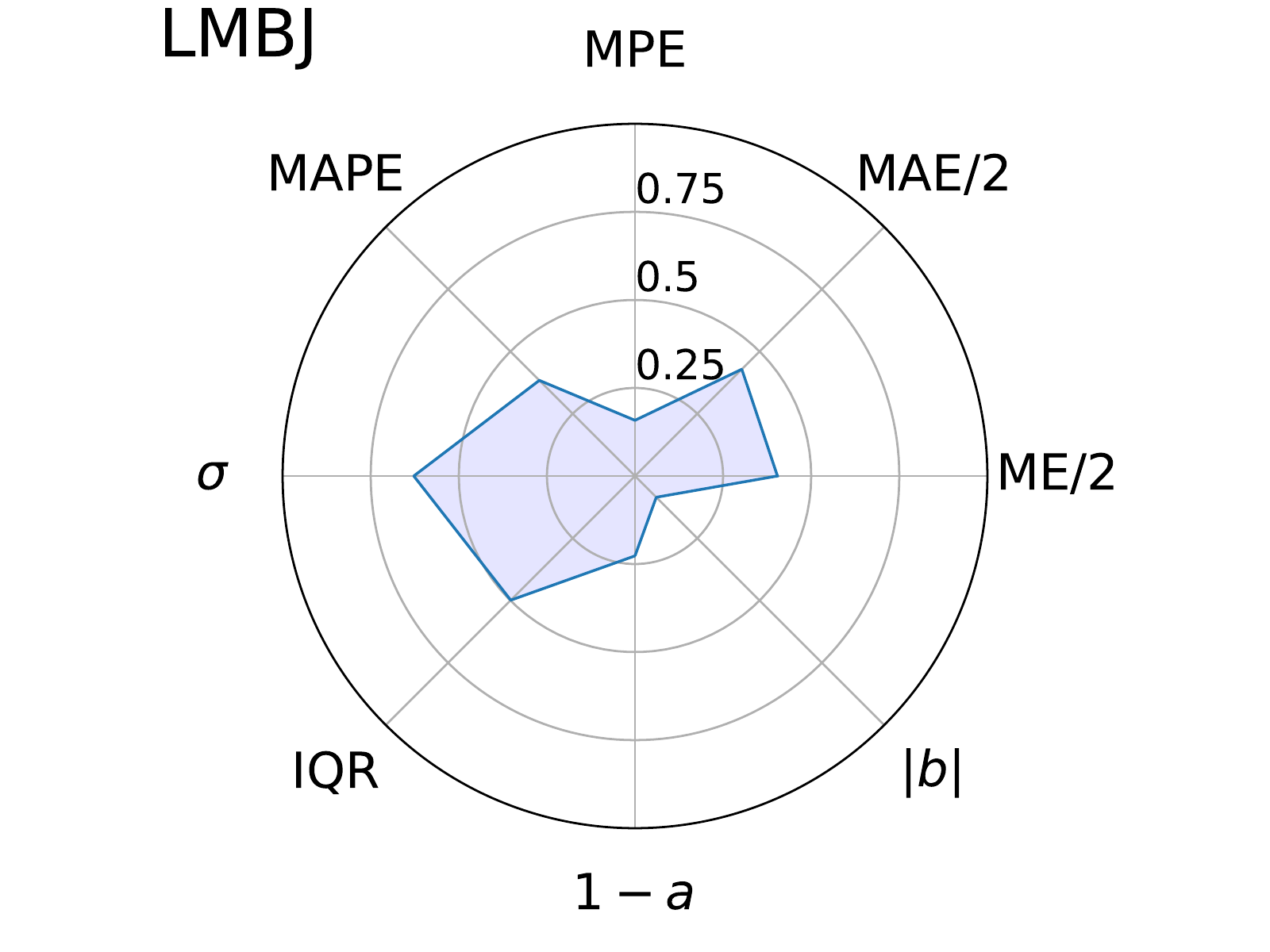}
  \caption{Radar charts of the statistical quantities in Tab.~\ref{tab:results} for HSE06 (C2DB), PBE, and LMBJ.}
  \label{fig:fig5}
\end{figure}

Having identified the optimal values of the LMBJ parameters, $\sigma= $\unit[4]{\AA} and $r^{\mathrm{th}}_s =$ \unit[5]{bohr}, we used them to calculate the band gaps of all 276 materials in the control data set. We present the graphical comparison of the results in Fig.~\ref{fig:fig4} and give various statistical quantities such as the mean error (ME $=\sum^{n}_{i} (y_i-y_{i,GW})/n$), mean absolute error (MAE $=\sum^{n}_{i} \left|y_i-y_{i,GW}\right|/n$), mean percentage error (MPE $=\sum^{n}_{i} (y_i-y_{i,GW})/ny_{i,GW}$), mean absolute percentage error (MAPE $=\sum^{n}_{i} \left|y_i-y_{i,GW}\right|/ny_{i,GW}$), standard deviation ($\sigma=\sqrt{\sum^{n}_{i} (y_i-y_{i,GW}-\textrm{ME})^{2}/n}$), interquartile range (IQR), and the linear fit ($y = ax+b$) coefficients in Tab.~\ref{tab:results} as well as in Fig.~\ref{fig:fig5}. 
\begin{table*}[h!]
    \centering
    \begin{tabular}{llcccccccccc}
        set & XC & \# & false met. & ME & MAE & MPE & MAPE & $\sigma$ & IQR & $a$ & $b$  \\
        \hline
        all & HSE06 (C2DB) & 276 & 0 & 0.73 & 0.79 & 0.14 & 0.30 & 0.68 & 0.59 & 0.72 & 0.15 \\
        all & PBE & 276 & 0 & 1.48 & 1.48 & 0.42 & 0.53 & 0.96 & 0.93 & 0.57 & -0.11 \\
        all & LMBJ & 276 & 3 & 0.81 & 0.86 & 0.16 & 0.38 & 0.63 & 0.50 & 0.77 & -0.09 \\
        $sp$ & HSE06 (C2DB) & 161 & 0 & 0.90 & 0.90 & 0.25 & 0.25 & 0.60 & 0.56 & 0.73 & 0.05 \\
        $sp$ & PBE & 161 & 0 & 1.58 & 1.58 & 0.46 & 0.46 & 0.94 & 0.85 & 0.58 & -0.10 \\
        $sp$ & LMBJ & 161 & 0 & 0.87 & 0.89 & 0.23 & 0.26 & 0.62 & 0.56 & 0.75 & 0.02 \\
        $d$ & HSE06 (C2DB) & 115 & 0 & 0.48 & 0.63 & -0.01 & 0.36 & 0.70 & 0.67 & 0.74 & 0.23 \\
        $d$ & PBE & 115 & 0 & 1.33 & 1.34 & 0.37 & 0.64 & 0.98 & 1.10 & 0.55 & -0.11 \\
        $d$ & LMBJ & 115 & 3 & 0.72 & 0.81 & 0.06 & 0.56 & 0.63 & 0.38 & 0.81 & -0.20 \\
    \end{tabular}
    \caption{Statistical measures for calculations of band gaps of 2D materials with HSE06 (from C2DB), PBE, and LMBJ potentials for the data sets ``all materials'', ``$sp$ materials'', and ``$d$ materials''. All errors are calculated with respect to the $G_0W_0$ band gaps from C2DB. The statistical measures are, in order, the number of false metals, mean error (ME, in eV), mean absolute error (MAE, in eV), mean percentage error (MPE), mean absolute percentage error (MAPE), standard deviation ($\sigma$, in eV), interquartile range (IQR, in eV), linear fit ($y = ax+b$) coefficients.}
    \label{tab:results}
\end{table*}

Looking directly at the data set of ``all materials'', we immediately see that the LMBJ potential performs much better than PBE, and it gives results of essentially the same quality as HSE06. While the absolute error of PBE is \unit[1.48]{eV} (\unit[53]{\%}), for LMBJ it is only \unit[0.86]{eV} (\unit[38]{\%}), close to HSE06 with \unit[0.79]{eV} (\unit[30]{\%}). The deviation of the errors from the mean value ($\sigma$ and IQR) is in the case of LMBJ slightly better than with HSE06, and $\backsim$\unit[30]{\%} better than with PBE. Finally, performing a linear fit of the data leads to very similar parameters $a$ and $b$ for LMBJ and HSE06 (slightly better for LMBJ), both much closer to the ideal values $a=1$ and $b=0$ than PBE. These results can be also easily understood by viewing the radar charts in Fig.~\ref{fig:fig5} where the different statistical quantities are plotted on different radial axes. The blue area can be understood as a simple measure of the quality of the given XC potential, smaller area signifying better overall performance.

We obtain additional insight by observing the results for the two smaller sets ``$sp$ materials'' and ``$d$ materials'' (see Tab.~\ref{tab:results} as well as Tab.~1 and Figs.~1 and 2 of the Supplemental Material). As expected, the MAPE is in the case of $sp$ materials smaller than in the case of $d$ materials for all three XC potentials. Remarkably, for $sp$ materials, LMBJ performs equally well as HSE06 with MAPE(LMBJ) = \unit[26]{\%} and MAPE(HSE06) = \unit[25]{\%}, while PBE gives much larger errors (MAPE = \unit[46]{\%}). In the case of $d$ materials, the quality of LMBJ (MAPE = \unit[56]{\%}) lies between HSE06 (MAPE = \unit[36]{\%}) and PBE (MAPE = \unit[64]{\%}). Here, we note a different behavior for $d$ materials with $E_{G_0W_0} \in (0,3)$ eV and $E_{G_0W_0} \in (0,3)$ eV, as shown in Tab.~1 and Fig.~2 of the Supplemental Material. In the former case, LMBJ (MAPE = \unit[68]{\%}) does not improve over PBE (MAPE = \unit[69]{\%}) and remains much worse than HSE06 (MAPE = \unit[40]{\%}), while in the latter, LMBJ is again as good as HSE06 (MAPE = \unit[22]{\%} for both), reducing the error of PBE by more than \unit[50]{\%} (MAPE = \unit[88]{\%}). Note, however, that part of this error might also be related to the difficulty of standard $G_0W_0$ to describe $d$-electron systems. A larger set of experimental data is therefore necessary to draw definitive conclusions. 

\section{Conclusions}
In summary, we extracted a set of 298 stable 2D non-magnetic materials from the C2DB database for which $G_0W_0$ band gaps are given. We chose 22 of the materials distributed as evenly as possible over space groups, chemical compositions, and the band gaps. We then used this subset to optimize the parameters $\sigma$ and $r^{\mathrm{th}}_s$ of the LMBJ XC potential, obtaining \unit[4]{\AA} and \unit[5]{bohr}, respectively. We used the $G_0W_0$ band gaps as target quantity, due to the lack of sufficiently reliable experimental data for 2D materials.

We then evaluated the quality of the LMBJ XC potential for 2D materials by calculating the band gaps of the remaining 276 2D materials in the data set. Overall, the performance of the LMBJ potential is very close to the one of the HSE06 hybrid functional. More precisely, LMBJ performs as well as HSE06 in the case of $sp$ materials, while being slightly worse for $d$-materials. Furthermore, it outperforms consistently the standard PBE XC functional in both cases.
We emphasize, however, that the LMBJ calculations are orders of magnitude faster than HSE06 (and than $G_0W_0$). Therefore, LMBJ can be used for band gap calculations of 2D materials, especially for systems where the HSE06 or $G_0W_0$ are computationally too demanding, almost without loss of accuracy compared to HSE06. This makes accurate band gap calculations possible also for large 2D materials with many atoms in the unit cell.

\ack
This work was supported by the Deutsche Forschungsgemeinschaft (DFG, German Research Foundation) through the projects SFB-762 (project A11), SFB-1375 (project A02), MA 6787/1-1 and BO 4280/8-1. S.B. and T.R. acknowledge funding from the Volkswagen Stiftung (Momentum) through the project ``dandelion''. 

\section*{References}
\providecommand{\newblock}{}


\includepdf[pages=-]{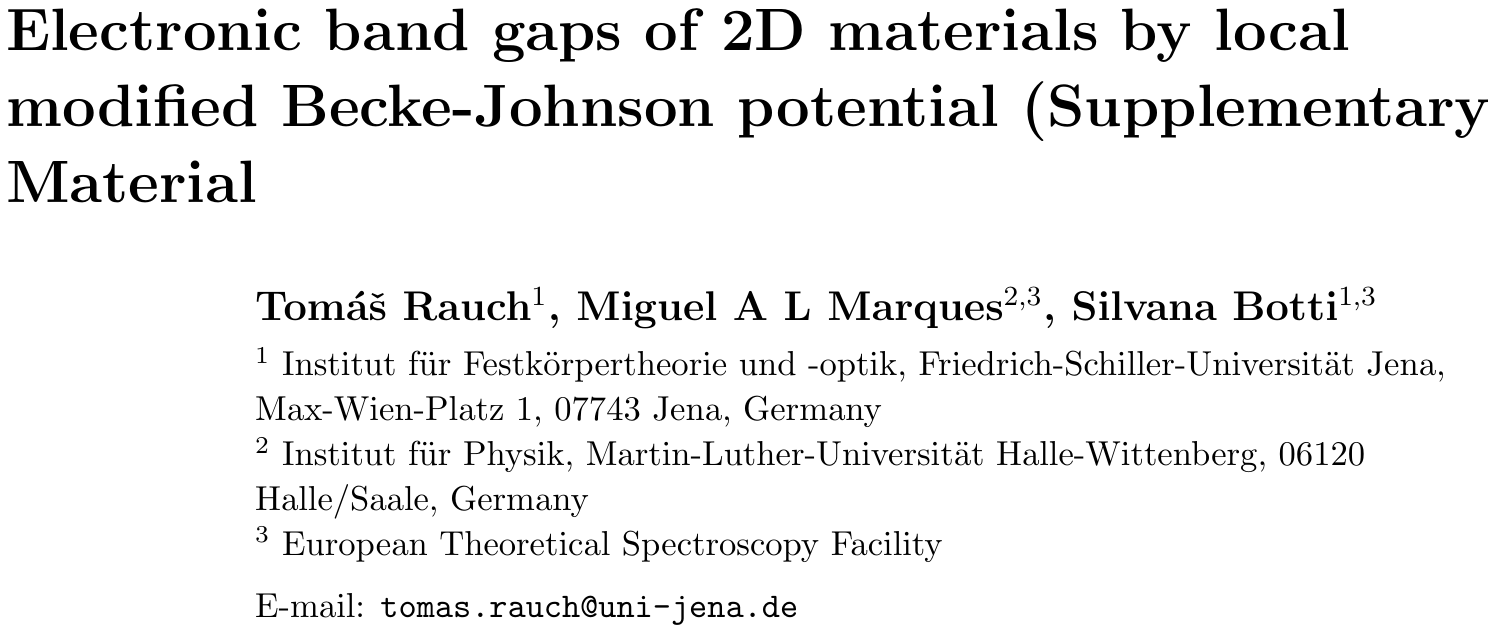}

\end{document}